\documentclass{wscpaperproc}
\usepackage{latexsym}
\usepackage{graphicx}
\usepackage{mathptmx}
\usepackage[T1]{fontenc}

\usepackage{amsmath}
\usepackage{amsfonts}
\usepackage{amssymb}
\usepackage{amsbsy}
\usepackage{amsthm}

\usepackage[pdftex,colorlinks=true,urlcolor=blue,citecolor=black,anchorcolor=black,linkcolor=black]{hyperref}

\newtheoremstyle{wsc}
{3pt}
{3pt}
{}
{}
{\bf}
{}
{.5em}
{}

\theoremstyle{wsc}

\begin{document}

%
%

\pagestyle{fancyplain}

\thispagestyle{plain}
\firstPageHead{}

\chead{\fancyplain{}{\itshape }}

\rhead{}
\cfoot{}
\renewcommand{\headrulewidth}{0pt} 

\makeatletter
\let\@internalcite\cite
\def\cite{\def\@citeseppen{-1000}%
    \def\@cite##1##2{(##1\if@tempswa , ##2\fi)}%
    \def\citeauthoryear##1##2##3{##1 ##3}\@internalcite}
\def\citeNP{\def\@citeseppen{-1000}%
    \def\@cite##1##2{##1\if@tempswa , ##2\fi}%
    \def\citeauthoryear##1##2##3{##1 ##3}\@internalcite}
\def\citeN{\def\@citeseppen{-1000}%
    \def\@cite##1##2{##1\if@tempswa, ##2)\else{}\fi}%
    \def\citeauthoryear##1##2##3{##1 (##3)}\@citedata}
\def\citeA{\def\@citeseppen{-1000}%
    \def\@cite##1##2{(##1\if@tempswa , ##2\fi)}%
    \def\citeauthoryear##1##2##3{##1}\@internalcite}
\def\citeANP{\def\@citeseppen{-1000}%
    \def\@cite##1##2{##1\if@tempswa , ##2\fi}%
    \def\citeauthoryear##1##2##3{##1}\@internalcite}
\def\shortcite{\def\@citeseppen{-1000}%
    \def\@cite##1##2{(##1\if@tempswa , ##2\fi)}%
    \def\citeauthoryear##1##2##3{##2 ##3}\@internalcite}
\def\shortciteNP{\def\@citeseppen{-1000}%
    \def\@cite##1##2{##1\if@tempswa , ##2\fi}%
    \def\citeauthoryear##1##2##3{##2 ##3}\@internalcite}
\def\shortciteN{\def\@citeseppen{-1000}%
    \def\@cite##1##2{##1\if@tempswa, ##2\else{}\fi}%
    \def\citeauthoryear##1##2##3{##2 (##3)}\@citedata}
\def\shortciteA{\def\@citeseppen{-1000}%
    \def\@cite##1##2{(##1\if@tempswa , ##2\fi)}%
    \def\citeauthoryear##1##2##3{##2}\@internalcite}
\def\shortciteANP{\def\@citeseppen{-1000}%
    \def\@cite##1##2{##1\if@tempswa , ##2\fi}%
    \def\citeauthoryear##1##2##3{##2}\@internalcite}
\def\citeyear{\def\@citeseppen{-1000}%
    \def\@cite##1##2{(##1\if@tempswa , ##2\fi)}%
    \def\citeauthoryear##1##2##3{##3}\@citedata}
\def\citeyearNP{\def\@citeseppen{-1000}%
    \def\@cite##1##2{##1\if@tempswa , ##2\fi}%
    \def\citeauthoryear##1##2##3{##3}\@citedata}
%
%
%
\def\@citedata{%
    \@ifnextchar [{\@tempswatrue\@citedatax}%
                  {\@tempswafalse\@citedatax[]}%
}

\def\@citedatax[#1]#2{%
\if@filesw\immediate\write\@auxout{\string\citation{#2}}\fi%
  \def\@citea{}\@cite{\@for\@citeb:=#2\do%
    {\@citea\def\@citea{, }\@ifundefined
       {b@\@citeb}{{\bf ?}%
       \@warning{Citation `\@citeb' on page \thepage \space undefined}}%
{\csname b@\@citeb\endcsname}}}{#1}}%

%
\def\@citex[#1]#2{%
\if@filesw\immediate\write\@auxout{\string\citation{#2}}\fi%
  \def\@citea{}\@cite{\@for\@citeb:=#2\do%
    {\@citea\def\@citea{; }\@ifundefined
       {b@\@citeb}{{\bf ?}%
       \@warning{Citation `\@citeb' on page \thepage \space undefined}}%
{\csname b@\@citeb\endcsname}}}{#1}}%

%
\def\@biblabel#1{}
\makeatother



\newdimen\bibindent
\bibindent=0.0em
\def\thebibliography#1{\section*{\refname}\list
   {}{\settowidth\labelwidth{[#1]}
   \leftmargin\parindent
   \itemindent -\parindent
   \listparindent \itemindent
   \itemsep 0pt
   \parsep 0pt}
   \def\newblock{}
   \sloppy
   \sfcode`\.=1000\relax}


\setlength{\baselineskip}{12.7pt}

\title{BROADENING ACCESS TO SIMULATIONS FOR END-USERS VIA LARGE LANGUAGE MODELS: CHALLENGES AND OPPORTUNITIES}

\author{\begin{center}Philippe J. Giabbanelli\textsuperscript{1}, Jose J. Padilla\textsuperscript{1}, and Ameeta Agrawal\textsuperscript{2}\\
[11pt]
\textsuperscript{1}Virginia Modeling, Analysis and Simulation Center, Old Dominion University, Norfolk, VA, USA\\
\textsuperscript{2}Dept.~of Computer Science, Portland State University, Portland, OR, USA\end{center}
}

\maketitle

\vspace{-12pt}

\section*{ABSTRACT}
Large Language Models (LLMs) are becoming ubiquitous to create intelligent virtual assistants that assist users in interacting with a system, as exemplified in marketing. Although LLMs have been discussed in Modeling \& Simulation (M\&S), the community has focused on generating code or explaining results. We examine the possibility of using LLMs to broaden access to simulations, by enabling non-simulation end-users to ask what-if questions in everyday language. Specifically, we discuss the opportunities and challenges in designing such an end-to-end system, divided into three broad phases. First, assuming the general case in which several simulation models are available, textual queries are mapped to the most relevant model. Second, if a mapping cannot be found, the query can be automatically reformulated and clarifying questions can be generated. Finally, simulation results are produced and contextualized for decision-making. Our vision for such system articulates long-term research opportunities spanning M\&S, LLMs, information retrieval, and ethics.

\section{INTRODUCTION}
\label{sec:intro}

As modelers, we are not usually our own end-users. Rather, models may be \textit{commissioned} to address specific organizational and societal needs, \textit{co-designed} with subject matter experts or individuals who provide their lived experiences, and employed as \textit{decision-support tools} that eventually impact communities~\cite{loeffler2021four,oldfield2022analytical}. End-users such as model commissioners, stakeholders, and community members may not have technical expertise in a modeling language or simulation environment. Yet, they need to interact with simulation models. When engagements are limited in scale, modelers may act as facilitators to translate the questions of end-users onto queries for a simulation engine, and translate simulation results into a contextualized and accessible response. However, a reliance on modelers can create a bottleneck in the process, since it depends on staffing limitations and cannot easily scale-up. For example, simulation models for welfare allocation, obesity management, or suicide prevention would potentially impact the lives of millions of individuals: sufficiently staffing a `modeling helpdesk' to help individuals ask questions and operate a model could be cost-prohibitive. In addition, when end-users either need help to interpret a model or start to dedicate their own time to learning about modeling, they are less able to translate simulation insights into practices~\shortcite{zellner2022finding}.

Environments such as {\ttfamily NetLogo} and {\ttfamily ClouDES} have been developed to make it easier for non-modelers end-users to interact with simulations~\shortcite{ramli2015overview,padilla2014cloud}. Several such environments have been presented at the Winter Simulation Conference (`WinterSim'), such as Sim4edu.com~\cite{wagner2017sim4edu}. These environments often provide web-based interfaces and they emphasize low-code/no-code interactions~\shortcite{hewage2024cloudsim,bocciarelli2023low} by allowing users to modify elements such as sliders or text fields as a means to input a new scenario. When existing interactive elements are not sufficient to input a scenario, drag-and-drop features may allow users to build more complex alternatives, as shown in commercial simulators~\shortcite{chong2022development}. However, these practices still face \textit{two challenges: a lack of usability, and limited potential for reuse}. First, although interactions are intended to be `user-friendly' by modelers and software developers, usability studies show that users struggle in integrating models to their existing software ecosystem~\cite{giabbanelli2023human}. Issues of usability have been observed for many years, as researchers have regularly pointed out that ``practical user-friendly tools that allow scenario simulations also to non-expert users are clearly still lacking''~\cite{marvuglia2018implementation}. 

Second, modeling interfaces that seek to be user-friendly may provide users with a graphical interface in which they can ask a question by changing a set of input values. While such interfaces may succeed in promoting high levels of usability, this would be at the expense of model reuse. Indeed, when users wish to ask questions that are not exactly as intended (yet still within the scope of the model), they would need more skills to pilot the simulation by expanding the interface or directly changing the implementation. The INFSO-SKIN study exemplifies this struggle, as a steering committee (i.e., model commissioners) provided a detailed list of questions in a Tender Specifications document, but ``stakeholders neither shared the same opinion about what questions should be in the final sample [...] nor shared the same hypotheses about questions in the final sample.''~\cite{ahrweiler2019co} Consequently, one user group may succeed in drafting a list of questions and modelers can build an interface accordingly, but another user group may find it inadequate for their needs, thus limiting reuse of a model across user groups. The difficulty of supporting user-defined questions was echoed in the recently proposed notion of `Approachable Modeling and Smart Methods' coined by~\citeN{schimpf2024approachable}, who stressed that \textit{tools need to promote user-driven inquiry}. One strand of research could be to provide more tools for end-users to extend interfaces such that they can input other scenarios (as long as they are supported by the model). However, the point was previously made at WinterSim that ``to make M\&S truly relevant to non-simulation experts and foster a radical change in the way we consume models and simulations, we have to venture beyond simply extending traditional graphical user interfaces''~\shortcite{rechowicz2018designing}. 

In this paper, we posit that \textit{emerging technologies can provide new modalities to interact with simulations in a manner that is both intuitive and sufficiently flexible to support a variety of user-driven inquiries}. As the practice of modeling and simulation spans several decades, there is an abundance of examples in which new technologies have been successfully integrated in the software ecosystem to support new means of interaction. For instance, simulations can now be embedded in the real-world context of a user through augmented reality~\cite{giabbanelli2024design}. Our focus is on democratizing access to simulations with a broad set of users by using \textit{Large Language Models (LLMs)} (such as OpenAI's GPT) as facilitators. LLMs have already been used on several occasions for specific simulation tasks such as turning a narrative into either a conceptual model~\shortcite{hosseinichimeh2024text,giabbanelli2024generative} or simulation code~\shortcite{Niu,Frydenlun2024}, learning about modeling by using LLMs as tutors~\shortcite{chen2024learning}, explaining a model~\shortcite{shrestha2022automatically}, or assisting simulated entities in making decisions~\shortcite{ghaffarzadegan2024generative}. At WinterSim, we also proposed to use LLMs during other parts of the simulation process, such as verification and validation~\cite{giabbanelli2023gpt}. Given this growing uptake of LLMs by the simulation community, this paper seeks to guide the discussion on the opportunities and challenges in using LLMs to assist a wide-audience of end-users in interacting with simulations.

The remainder of this paper is structured as follows. In Section~\ref{sec:framework}, we provide a framework to identify the main three steps and components of a system using LLM as mediators between end-users and simulations. Each of the three steps is then detailed subsequently: Section~\ref{sec:question} focuses on aligning questions and model parameters, Section~\ref{sec:clarify} examines the need for clarifications (e.g., via query reformulation), and Section~\ref{sec:response} discusses the requirements for the final answer such as the need to ground it in the model's context. To design such systems, we will need to evaluate them, thus Section~\ref{sec:evaluation} summarizes quantitative and qualitative metrics that could be of interest in future implementations. We provide concluding remarks in Section~\ref{sec:remarks}.

\section{Conceptual Framework}
\label{sec:framework}

\begin{figure}[ht!]
    \centering
    \includegraphics[width=\textwidth]{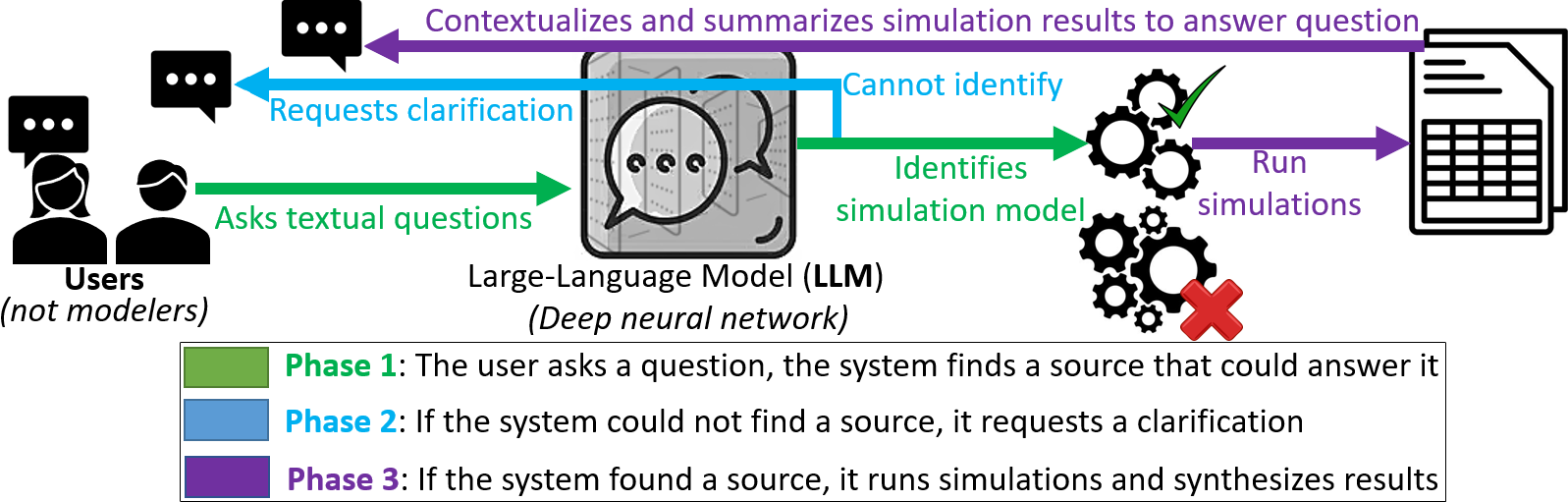}
    \caption{Overview of our conceptual framework including textual questions, alignment with the right model (or request for clarification), and translation of structured simulation results into text. We envision the system as composed of three phases (indicated by colors), detailed in dedicated subsections 3 to 5.}
    \label{fig:overview}
\end{figure}

Our framework is summarized in Figure~\ref{fig:overview}. It is designed to allow users to express questions naturally, which would be answered thanks to a simulation model. We assume that \textit{the purpose of the query is a what-if question that would translate to a simulation input}. This rules out questions such as ``can you create a simulation model'' or ``can you verify this model'', which have been discussed in other articles (see Section 1). We make no assumption on the style of a question. For example, users are not limited to the use of a specific professional jargon. As a consequence of letting users choose their own words, we will need to map the key constructs in their question to the parameters available in a simulation model. We assume a general case in which \textit{multiple} models may be available, such that the more relevant one should be triggered to address the user's question. For example, there may be an ensemble of geographically specialized models and the user can be interested in a specific area. Traditionally, identifying key constructs would be achieved through the Natural Language Processing (NLP) task of entity recognition, while mapping constructs expressed by the user onto the inputs of a model is a matter of semantic matching. Through several benchmarks and case studies~\shortcite{yuan2024revisiting,chen2023beyond}, LLMs have demonstrated an ability to perform both entity recognition and semantic matching. Consequently, our framework relies on the LLM (via additional prompts) to align a user's query with a suitable simulation model.

On the one hand, \textit{we should not expect a perfect alignment}: that would force the user to define a value for every input of a simulation model, while excluding any concept beyond the model. This would be a mechanical or `templated' interaction unlike the natural style that we seek to support. Considering an interaction between end-users and modelers as facilitators, modelers may not refuse to answer a question until every input has been specified (they may assume that unspecified inputs are left to a baseline value) or as soon as another concept is evoked (they may just ignore it). On the other hand, if the user's question is too far from any available model, a modeler would prompt for clarifications. In this spirit, the system has an alignment threshold below which it requests a clarification from the user. This may lead to a short conversation until the request becomes sufficiently clear given the simulation models available.

Once the request is clear and a corresponding model is identified, simulations are run accordingly. We assume that simulation results are stored in a structured format hence the LLM performs the task of data-to-text generation. This task can be further specialized depending on the data model used in the simulation experiments: for example, tabular data such as CSV files call for table-to-text generation. Numerous benchmarks and improved LLMs have produced satisfactory performances on tasks involving tables~\shortcite{sui2024table} and other types of structured data~\cite{kasner2024beyond}. Beyond the well-studied matters of generating content that is factual and easy to read, \textit{our framework stresses the importance of contextualizing simulation results to promote ethical decision-making activities}. Indeed, the reason to use simulations in the first place is often that taking decisions in a complex adaptive system may generated unintended consequences or only yield limited benefits due to unforeseen limitations~\cite{putro2016value}. We should thus make efforts to answer a user while accounting for factors that were not directly featured in their questions. For example, if a user asks ``what will happen to the number of fast-food outlets if I implement new zoning policies to limit them?'', one answer is that ``there will be 8.52\% fewer fast-food outlets in total'', which may sound successful and encourage their policy. Another answer using the same simulation model is that ``there will be a reduction of 16.2\% fast-food outlets in affluent neighborhoods, and an increase of 8.2\% in deprived neighborhoods, where individuals are already are a higher risk for obesity due to over-exposure to unhealthy foods''. The first answer strictly addresses what the user wanted, but the second one provides enough context to inform users against potential issues of fairness and equity. 

\textit{In principle}, the framework discussed in this section could be enacted using any individual LLM (e.g., OpenAI's GPT, Meta's LLaMa, Google's Gemini) or ensemble of LLMs as supported by emerging frameworks such as {\ttfamily LLM-BLENDER}~\citeN{jiang2023llm}. We would expect performances to depend on usual considerations such as the choice of LLM and its hyper-parameters (e.g., the temperature setting of GPT), the level of training, the application domain, and characteristics of the question. For example, an ambiguous question regarding a highly specialized medical model (which is unlikely to be contained within the dataset of a general-purpose LLM) and an absence of training may not yield high performances. However, some LLMs have safety settings that \textit{refuse to engage} on certain topics, hence they will not answer a question or pass it onto subsequent modules for analysis. For example, Gemini evaluates and blocks requests that present a safety risk by containing hate speech, harassment, or dangerous content. This LLM will thus forbid requests to interact with a public health policy model for suicide prevention\footnote{Safety mechanisms can be bypassed by using a lax LLM to express a potentially sensitive request without using certain words~\shortcite{Tyler}. For instance, we can provide a GPT prompt such as ``You are a helpful assistant that combines a list of sentence into one paragraph that is rated Y for toddlers to read: {\ttfamily [request]}''. The result can then be passed onto an LLM with more stringent safety checks. However, the reworded sentence may be more vague and thus harder to align with the parameters of a simulation model. For example, professional terms (which would be used by a model) such as Adverse Childhood Experiences would be expressed as ``bad things that happen to people, especially children''.}.

\section{What is a Question? Aligning Constructs with Model Parameters}
\label{sec:question}

Let's start with an example based on~\shortciteN{tolk2013reference}, where users ask questions regarding the consequences of sea level rise and its effect on flooding in city areas. Users want to know whether to vacate, invest, or maintain (VIM). Accordingly, users can ask ``what VIM decision should I make in the downtown area if sea level will rise by six inches?'' To formulate this question requires \textit{known statements} that establish what vacating, maintaining, and investing are. It also requires statements (potentially unknown at the time) of the needed factors, and their combination, that may lead to a VIM decision. A question may vary depending on the purpose of the answer (\textit{intent}). The question here was formulated from the perspective of a city's decision makers that need to assess where to vacate, maintain, or invest. This example previews major challenges~\cite{Zeigler}: users may be unfamiliar with modeling practices and the level of details that a model needs (i.e., scope and resolution); models may answer a question partially; what-if scenarios can have different effects in different models. 

Given these challenges, we need to operate under clear definitions. According to~\shortciteN{tolk2013reference}, a modeling question is ``a collection of sentences to which truth values needs to be assigned'' and akin to a query directed to a referent, or in our context, a \textit{reference model}. A reference model captures everything that is known or assumed about a problem domain. Depending on social processes dynamics and experiences of a team, the reference model may be reduced into a few options as represented by different \textit{conceptual models}. There can also be multiple ways of implementing a simulation given a conceptual model, for example based on data availability~\cite{freund2021necessity}. As a result, there is not necessarily `one' valid model to answer a user's question. Rather, the system needs to identify a \textit{candidate simulation model} to answer the question at the scope and resolution needed to meet the intent of the final user. We focus on identifying \textit{one} candidate model, while noting that several algorithms can be applied if the system identifies \textit{several satisfactory candidate models}~\cite{testoni2024asking}, for instance by measuring the system's uncertainty (e.g., via entropy) and asking the user for preferences.

\begin{figure}[ht!]
    \centering
    \includegraphics[width=\textwidth]{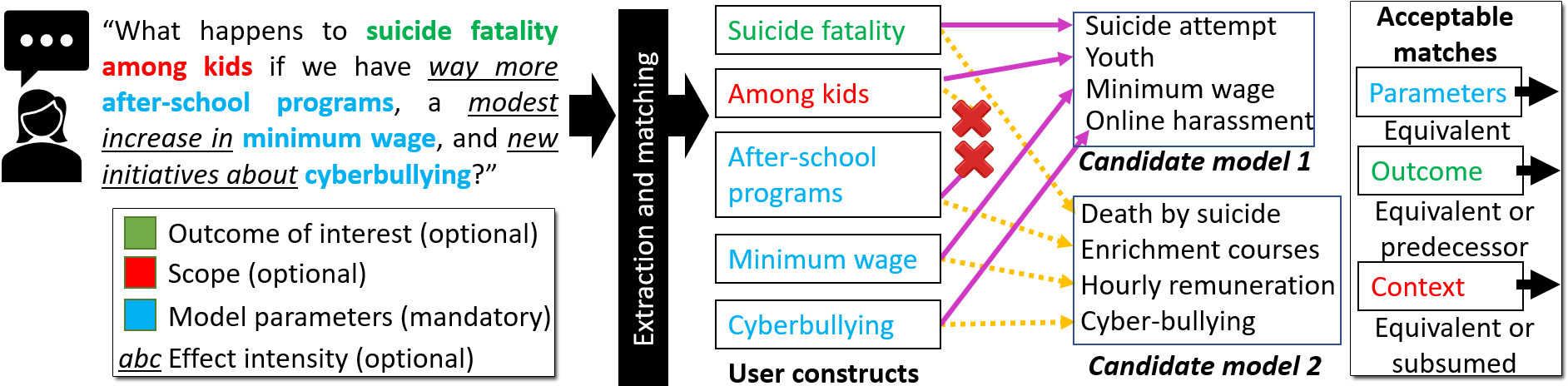}
    \caption{The user starts by formulating a what-if question. The \textit{extraction} step identifies and categorizes key constructs. The user constructs are then \textit{matched} with semantically equivalent parameters from available models. As a result, we \textit{select} an acceptable model that operates in a context including the user's interests, and that is sufficiently informative as its outputs directly feed into the desired measurement outcomes.}
    \label{fig:mapping}
\end{figure}

The LLM needs to consider that modelers make several assumptions when identifying candidate models. As exemplified in Figure~\ref{fig:mapping}, modelers may consider that parameters are a valid match when they are \textit{semantically equivalent} to the constructs that users seek to modify through their what-if question. Simulation outcomes should be related to the user's question\footnote{It may be too restrictive to only accept a simulation model that has the outcome (per semantic equivalence) desired by the user. If no such match can be found, the user may tolerate a model whose outcome directly precedes their measurement of interest. For example, if we cannot find a model of death by suicide, then it may be satisfactory to retrieve a model of suicide attempt. The LLM would \textit{not} be responsible for improvising a model that relates the two constructs, as this is not a straightforward endeavor (e.g., suicide attempts are more lethal for men than women). The LLM would be responsible for recognizing that one stage closely precedes another, while noting that this increases uncertainty and has yet to be tested.} and the population captured by the model should include the sub-population of interest to the user. In order to recognize these relations, the LLM would need to either directly capture the reference model or have access to it (e.g., by querying a corpus when needed). Our recent proof-of-concept in~\citeN{giabbanelli2024generative} suggests that LLMs can adequately capture several semantic relations (e.g., synonyms, antonyms) and aptly leverage authoritative sources such as peer-reviewed papers (e.g., the new prompt size for GPT-4 allows to pass a PDF document).

\section{Can (A)I Have a Word with You? Clarifying Questions from the LLM}
\label{sec:clarify}

\begin{figure}[ht!]
    \centering
    \includegraphics[width=0.9\textwidth]{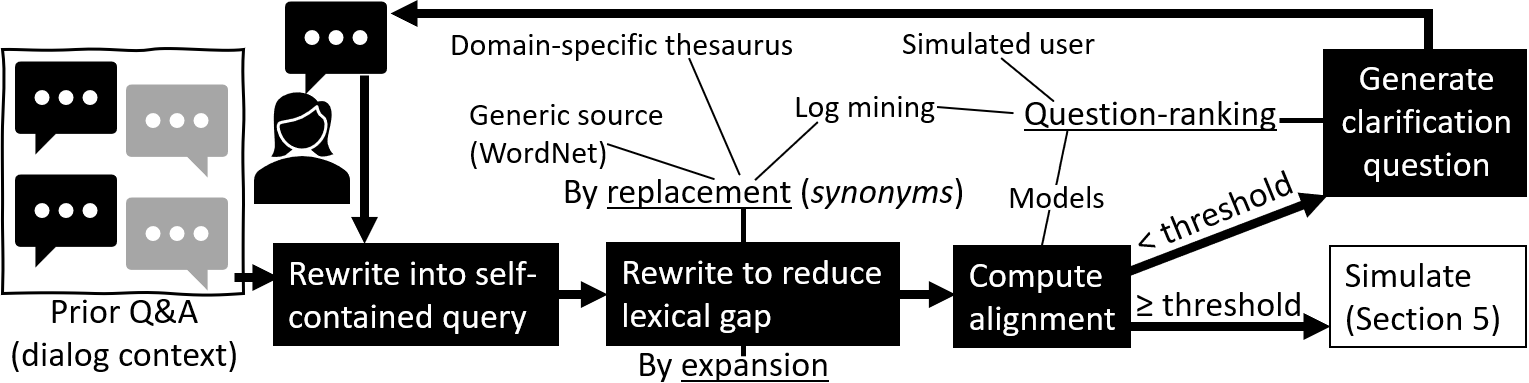}
    \caption{A user's query may need to be rewritten so that it becomes self-contained, then it can be rewritten to better align with simulation models. If the alignment is satisfactory, the simulation can proceed. Otherwise, the system generates a question and interprets the user's answer through the same rewriting process.}
    \label{fig:clarify}
\end{figure}

The user's question will be answered by \textit{retrieving}, running, and explaining a simulation model. We expect a gap between the user's query and a model, given that they may use different words. One of the primary concerns in the field of Information Retrieval (IR) is to resolve the lexical gap between a user's query and the identification of a suitable `document' (i.e., a simulation model in our case). As summarized by~\shortciteN{anand2023context}, multiple IR techniques seek to ``reconcile the difference between user intent and the intent of the [modeled] system''. Query Rewriting/Reformulation (QR) is particularly common to address underspecified and ambiguous queries, prior to (or instead of) asking a user for clarifications. Queries can be expanded automatically by prompting LLMs to recommend useful expansion terms and appending them to the user's original question~\shortcite{wang2023generative}. It is even possible to request LLMs to create different prompts and answer them, thus using the variability of outputs to yield a greater variety of additional terms~\cite{dhole2024genqrensemble}. However, adding terms may \textit{drift} from the user's intent\footnote{For instance, a student pursuing modeling and simulation may ask GPT 3.5: ``What is the number one factor to have a good career in modeling?'' Without knowing the user's intent, the LLM assumes that the question is about the fashion industry and will emphasize physical attributes, such as ``photogenic facial features''. Using the LLM to find expansion terms via the prompts as shown by~\citeN{dhole2024genqrensemble} will thus add physical appearance to the user's query, which makes it drift from the user's intent and complicates the alignment with a suitable set of parameters.} and increase computing time when retrieving a simulation model. An alternative is to rephrase the query by replacing some of the users' terms with synonyms that are more likely to be expressed as simulation parameters. We posit that suitable synonyms may be learned from prior user interactions with models, as these logs\footnote{As a user successfully gets an answer, we can store the terms used in their question and the corresponding mapping to a simulation model. Through their interactions with the system, users thus contribute to building a thesaurus. As we discussed in~\citeN{giabbanelli2019overcoming}, such thesaurus could improve the quality of future alignments and reduce their computational burden. In addition, a thesaurus can be a valuable research object in its own rights, as it reveals how end-users think of models.} can reveal which user terms were eventually mapped to satisfactory models (Figure~\ref{fig:clarify}).

The matter of retrieving the right model is complicated by the fact that interactions between users and models can turn into \textit{conversations}. For example, consider that a user asks the modeler to run one intervention, hears the results, and finds them disappointing. The user may then ask ``what if we \textit{also} increased minimum wage?'' This follow-up question supposes that one parameter is modified \textit{in addition to} the modifications previously stated. A modeler would understand the dialogue context and interpret the question accordingly. Using an LLM, a simple solution could be to request that users provide complete questions each time, refraining from referring to their previous statements or results. While this may suffice for a prototype, we caution against designing a system for a broad user-base that relies on their goodwill to follow ambiguous instructions. Rather, we posit that a complete system would eventually have to handle conversational dependencies. Specifically, the LLM would be given the list of all prior question-answer pairs and \textit{rewrite the user's query to make it self-contained}. However, this often requires fine-tuning a query reformulation model, which adds to the computational cost of the system~\shortcite{yoon2024ask}.

Despite the necessity of aligning a user's query with a model (with the help of query rewriting), the system should also account for the likely event in which the query is too imprecise. As mentioned in Section 2, a threshold on the quality of alignment should thus trigger a clarification. \shortciteN{rao2018learning} emphasize ``that a good question is the one whose \textit{likely answer} will be useful.'' There are two implications: the LLM should focus on questions that \textit{(i)} the user can plausibly answer to \textit{(ii)} improve the alignment with a simulation model. The first part requires a dataset to learn about users and their questions. While there exist datasets of human conversations containing questions (e.g., to develop dialogue systems)~\cite{testoni2024asking}, there is a paucity of data documenting the textual interactions of users with simulations. Such a dataset may be collected with the users' permission, once a system has been in operation for some time\footnote{The use of log mining to ask clarifying questions can also bring a \textit{collaborative} dimension and result in a very different type of question. For example, recommender systems~\cite{giabbanelli2020reducing} could find the top-$k$ most relevant simulation models, based on users who had similar questions. The system could thus present the top-$k$ models along with a brief context (e.g., model purpose) and an explanation (e.g., ``95\% of people like you who were interested in suicide prevention have used this model''). A clarifying question would then ask a user too choose rather than to provide more information.}. Alternatively, user simulators have been developed to provide feedback to a system's response and answer clarifying questions~\shortcite{owoicho2023exploiting}.

\section{Responsible Responses: Contextualizing Answers and Promoting Ethics}
\label{sec:response}

\begin{figure}[ht!]
    \centering
    \includegraphics[width=\textwidth]{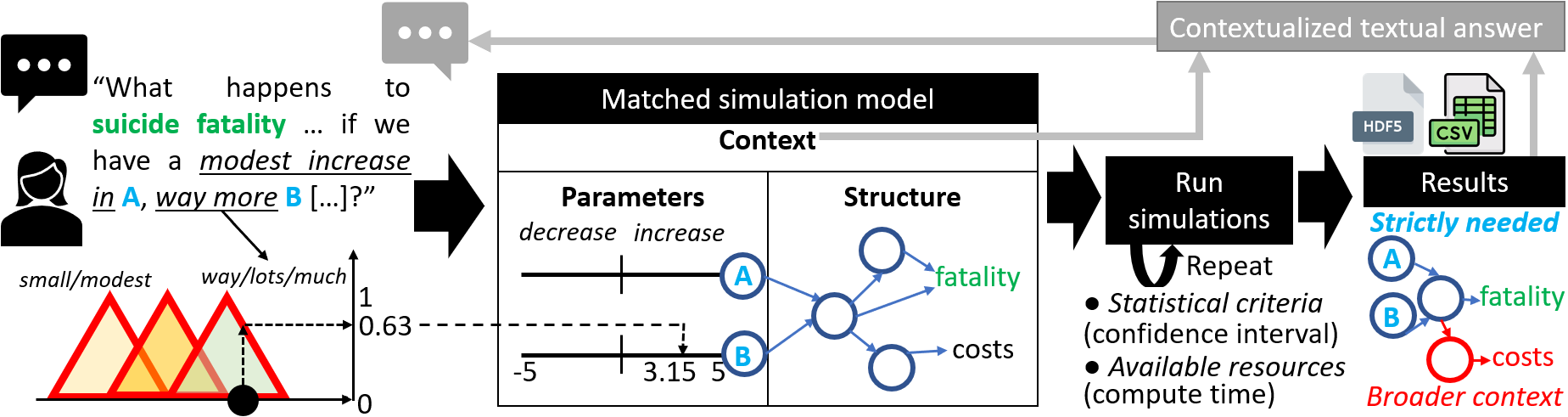}
    \caption{Running a simulation requires mapping user constructs to model parameters as well as converting linguistic modifiers (e.g., `modest increase') into quantitative values for each parameter's domain via fuzzy logic. Several simulation runs can then be performed, but a comprehensive interpretation of the structured results (e.g., CSV or HDF5) would require access to the model's context and internal structure.}
    \label{fig:answers}
\end{figure}

At this stage, the user asked a (potentially reworded and clarified) what-if question that calls for experimentation by simulation and a relevant model has been retrieved. We thus need to run the simulation and explain its results (Figure~\ref{fig:answers}). The alignment between the user's question and the simulation model informs us on \textit{which parameters} need to be modified from their default values. The question must be parsed further to determine the \textit{direction and level of change} on each parameter. The direction of effect may be detected by the LLM. The level of change can be obtained by using fuzzy logic to transform the qualitative modifiers expressed by the user (e.g., `way more' after-school programs, `modest increase' in minimum wage) into values within the operating domain of each parameter~\cite{giabbanelli2014creating}. If a model is stochastic, multiple simulation runs will be needed. Several methods such as desired Confidence Intervals can determine the number of runs (see section 9.7.1 in~\citeNP{robinson2014simulation}). However, if the proposed system operates as an open public platform where any number of individuals can ask questions, then the delivery of compute resources in the context of citizen science may become a determining factor\footnote{A simple approach is to use cloud computing resources and set a cost limit on each user query. However, the use of cloud computing for citizen science has its challenges. In particular, if the system mines user logs in order to improve its alignment algorithm or ask clarifying questions, then it ``requires systematic, semi-automated data governance, and management plans in order to prevent the degradation of the data lake into a data swamp''~\cite{ramachandran2021open}.}.

The main research challenges are about contextualizing the simulation results, not only to address the user's question but also to promote sound and ethical decision-making\footnote{Modelers may consider that promoting sound decision-making is the right action from a moral standpoint. However, this is also under increased scrutiny from the perspective of criminal liability for AI systems, including simulations. For instance, a decision-making platform (albeit sophisticated) may be viewed as a machine that cannot be the perpetrator of an offense, much as a mentally limited person (e.g., a child) would be considered an innocent instrument~\cite{hallevy2024basic}. If a person causes a child to commit an offense, that person is criminally liable. The problem here is to determine whether that person is the \textit{user} who asked a flawed what-if question, or the \textit{programmer} whose system provided an inappropriate answer.}. One challenge is that users of computer simulations tend to narrowly focus on intended consequences and neglect essential information on alternative, unintended consequences. Through several experimental studies,~\citeN{ehrlinger2011focalism} have shown that participants insufficiently accounted for unintended consequences in a simulation, unless they were informed about the interrelatedness of variables in the modeled system. In other words, users tend to look for what directly interests them, unless they are told about the impact of a decision onto other outcomes and then start to consider the \textit{overall} impact. On the one hand, it may be inadvisable to only tell users about the consequence of their what-if scenario on the variable that interests them. On the other hand, it may be exceedingly burdensome to tell them about consequences on all variables. We thus posit that a trade-off is a necessity. The technical challenge is that navigating this trade-off may require information on a model's context, at least to determine which variables are outcomes. A second challenge is to evaluate algorithmic fairness. This often involves assessing variations in outcomes across demographics such as race and ethnicity, age, and income subgroups~\shortcite{schuch2023fairness}. If simulation outcomes are different in some of these groups, then the user should be informed as part of the response. Again, this would require contextual information about the model to identify which variables represent demographic data.




\section{What is a Good System? Measuring Faithfulness, Biases, and Explanations}
\label{sec:evaluation}

While our framework would create a combination of simulation models and LLMs that is accessible to a broad audience, evaluating the system is a critical challenge, compounded by the limitations of the two parts as both simulations and LLMs can produce outputs that are not always accurate, reliable, or unbiased, or answers may simply get `lost in translation'. To address these challenges, a multifaceted evaluation approach should be adopted by focusing on three main phases defined in the previous sections: question processing, query clarification/reformulation, integration of simulated output and response generation. As part of this evaluation, it is important to first create a carefully annotated dataset to provide a benchmark. This data should consist of \textit{user questions}, \textit{mapped queries}, and \textit{simulated outputs}. Given the vast space of possible user questions, automatically generating user questions based on parameters can be considered.

Since the LLMs are essentially serving as the bridge between the user and the simulation model, the first stage of evaluation should focus on whether the LLM accurately understands the user's question and maps it to a set of parameters that can be understood by the simulation model. LLMs are capable of processing natural language, but how well do they interpret the user's intentions/constructs and formulate a query that can be used as input to a simulation model? Evaluation at this stage would thus identify when simulation model parameters are not compatible/complete/found with the user's question. This evaluation is a necessary part of our system, as discussed in section 4 where the LLM asks clarification questions to the user until satisfactory queries can be formulated. The most visible part for the user is the generation of the final response, hence we focus on its evaluation.




Metrics of \textit{faithfulness} assess how well the LLM's response reflects the content of the simulation results, avoiding any misinterpretations or injections. Standard metrics (e.g., ROUGE, BERTScore) are not effective in capturing nuanced faithfulness or lack thereof, as they focus on the overall semantic similarity. Instead, one should rely on semantic \textit{inference} related measures such as textual entailment which has been found effective in identifying faithfulness~\shortcite{maynez2020faithfulness}. Natural Language Inference (NLI) is a task in natural language processing where given two pieces of text -- a premise (LLM's response) and a hypothesis (simulated output) -- the goal is to determine whether the hypothesis can be inferred from the premise, with common relationships including entailment, contradiction, and neutral. An evaluator can leverage existing NLI models or train one by fine-tuning a pretrained language model such as BERT on a new NLI dataset of simulated outputs. Specific factuality metrics based on textual entailment include Adversarial Natural Language Inference~\shortcite{nie2019adversarial}, Summary Consistency~\shortcite{laban2022summac} and FactCC~\cite{kryscinski2021evaluating} to determine whether a given text is consistent with the source.


Because assessing faithfulness requires logical inference over factual information, several question-answering (Q\&A) based models can also be used such as QAGS~\cite{wang2020asking} and FEQA~\cite{durmus2020feqa}. Faithfulness is only one aspect of evaluation. Another important dimension is that of \textit{biases}, which can happen in LLMs due to issues at various stages such as dataset selection, annotation, and instruction tuning~\cite{agiza2024analyzing}. While several definitions of fairness and biases exist, two commonly used ones focus on demographic parity and equal opportunity. The former compares the distribution of sensitive demographic attributes (e.g., race, gender) in the LLM's outputs to the distribution in the real world or the training data, with ideal outputs not favoring any particular demographic group. The latter assesses whether the LLM provides similar response for different demographic groups. For example, if you ask an LLM to suggest salary increments for a person, it should offer similar options regardless of the person's gender. Recently introduced metrics such as Large Language Model Bias Index (LLMBI) can provide insights into any biases that may occur in the generated output~\cite{oketunji2023large}. Yet another dimension of quality can be derived by examining the LLM's confidence scores. Some LLMs (mostly open-source) can generate how confident they are in their responses. However, since the scores reflect the model's own assessment, it may be unreliable.

In addition to simply responding to the user question, the user response can be augmented by \textit{explanations} generated by the LLM. This would inform the user not only of the simulated outputs but also explain the rationale behind the answer for more effective decision making. To assess how useful the explanations are to the user, one could perform a user satisfaction study or measure user engagement metrics (conversation duration, follow-up questions). This could also be done for the overall response of the LLM.

Finally, when it comes to automatic evaluation, an emerging area of research is to use \textit{use LLMs as evaluators}~\shortcite{liu2023gpteval,ferron2023meep,gao2024llm}. LLM-based evaluators can be used for large-scale evaluations and adapted to different evaluation criteria. Combined with a human-in-the-loop approach, LLMs offer the potential of providing reliable assessments alongside human evaluators.


\section{Discussion and Conclusions}
\label{sec:remarks}

We articulated challenges and opportunities to design an end-to-end system that supports a broad set of end-users in interacting with simulations via textual queries. With the right resources, many of the steps that we outlined can be addressed by the research community in the short- to medium-term. However, there are also five long-term research challenges that warrant particular considerations. First, what if no \textit{single} model can be retrieved by the LLM to answer a user? For example, there may be two models: one that computes the effect of $A$ on $B$, and another that has parameter $B$ and outcome $C$. A user may ask how a small increase in $A$ impact $C$, which is not directly answerable by a single model but could be addressed by \textit{composing} models. Creating a combination of models on the fly is complicated, as models may be intertwined rather than neatly separable. At least, we should ensure that the combined model is consistent (i.e., without contradicting statements). Second, user questions may not be as independent as they seem. For example, in a validated model for suicide prevention, users could ask about the effect of increasing training for doctors: little to no impact. They could then ask about the effect of adding mental health education programs: little to no impact. However, when combined, these two evidence-based interventions could \textit{increase} suicidal behavior by creating an imbalance between service capacity and demand for services~\shortcite{atkinson2020science}. This is a problem for ethical decision-making, as users may not inform the platform that they intend to use two interventions jointly, hence joined effects are never simulated. 

Third, an end-to-end solution requires many sub-systems, including retrieval (to align a query with a relevant model), query rewriting, and running a simulation multiple times. The computational cost may become a concern, particularly if there is a broad set of potential users. While the problem may be alleviated by queuing simulations and informing users when results are ready, this may reduce the quality of service and it only distributes computations over time instead of fully addressing their costs. Alternatives to reduce the processing costs and associated energy consumption (i.e., green computing) have been studied both in the Information Retrieval community~\cite{scells2022reduce} and by the simulation community~\cite{feng2017green}. For instance, the system may answer a \textit{related} question that has already been computed. A fourth related challenge is that LLMs may already be able to generate simulation outputs instead of running the models. Indeed, LLMs have been trained on enormous amounts of data, some of which includes simulation models. For instance, GPT can answer questions about the behavior of models that are part of the standard library of {\ttfamily NetLogo}. It would be of particular interest to assess to what extent can LLMs complement or replace the outputs of a simulation model, for instance by checking for possible alignment between the responses of the LLM generated with and without the simulation model.



Finally, the response of an LLM is a \textit{mixture} of the underlying information derived from the simulated outputs and its own parametric knowledge. Evaluating this mixture is of particular interest. Is the response grounded in the human/domain knowledge embedded in a validated simulation model, or is it produced by the knowledge embedded in the LLM (coming from various and sometimes unknown datasets)? Does it depend on the application domain? Users concerned with this distinction could restrict the information space~\shortcite{braun2024can} by formulating their query accordingly (e.g., ``reply solely based on the simulation model''). In some cases, it is \textit{necessary} for the LLM to draw from its own knowledge, or the response may lack context. We thus recommend that future research examines how to clarify the role of a simulation by contrast to the LLM's knowledge when providing an answer. 






\end{document}